\documentstyle[array,preprint,prl,aps]{revtex}
\draft
\newcommand{\be}{\begin{equation}}
\newcommand{\ee}{\end{equation}}
\newcommand{\ben}{\begin{eqnarray}}
\newcommand{\een}{\end{eqnarray}}

\begin{document}
\draft

%@twocolumnfalse\endcsname

\title{How fundamental is the character of thermal uncertainty relations?}

\author{ ~F. ~Pennini\thanks{e-mail: pennini@venus.fisica.unlp.edu.ar},
A.~ Plastino\thanks{Corresponding Author, e-mail:
plastino@venus.fisica.unlp.edu.ar}, and A.~ R.~ Plastino\thanks{
e-mail: plastino@sinectis.com.ar}}
\address{National
University La Plata (UNLP) $\&$  Argentine National Research
Council (CONICET)
\\ C.C. 727, 1900 La Plata, Argentina }

\maketitle

\begin{abstract}
We show that thermodynamic uncertainties {\it do not preserve
their form} if the underlying probability distribution is
transformed into an escort one. Heisenberg's relations, on the
other hand, are not affected by such transformation. We conclude
therefore that the former uncertainty cannot be as fundamental as
the quantum one.


\end{abstract}
\maketitle
%============================================================================
\newpage
\section{\strut Introduction}

%============================================================================
Thermodynamics ``uncertainty'' relations have been the subject of
much interesting work over the years (see, for instance,
\cite{ro,Mandelbrot,lavenda}). An excellent, recent review is that
of Uffink $\&$ van Lith \cite{Uffink}. We will  be interested here
in these uncertainty relations insofar as they are derived by
recourse to statistical inference \cite{roybook}, with emphasis
upon Mandelbrot's results \cite{Mandelbrot}.

Heisenberg's uncertainty relations and Bohr's complementarity
principle constitute two pillars of $20$-$th$ century science.
These two prominent authors have suggested that there is a
classical  analogue of the complementarity principle, specifically
between temperature and energy \cite{promine}. Although such ideas
have not received  general acceptation, several renowned authors
have defended them, as exemplified by, among others, Refs.
\cite{ro,Mandelbrot,lavenda}. These claims remain still
controversial (see  \cite{Uffink,physto1,physto2}).

 We wish in this Communication to add a footnote to the
  controversy {\it by focusing attention upon particular
   aspects of the  thermal
uncertainty derivation of Mandelbrot's} \cite{Mandelbrot}. This
derivation contains as an essential ingredient the information
measure introduced  by Fisher in the twenties \cite{roybook,f7}.

Mandelbrot \cite{Mandelbrot} is one of the first authors that
linked statistical physics with the theory of statistical
inference, adopting the viewpoint that one can work in statistical
mechanics directly with probability distributions over macroscopic
variables, the microscopic substructure (e.g., phase space) being
largely superfluous. Let $U$ denote the internal energy.
Mandelbrot \cite{Mandelbrot} established which is the form of the
probability distribution $p_\gamma (U)$ that allows for an
adequate description of the energy  fluctuations of a system in
contact with a heat bath at the (inverse) temperature
$\gamma=1/T$. The ensuing distribution  turns out to be the
celebrated, text-book (Gibbs') canonical one \cite{gibbs}, an
exponential probability density. A quite interesting uncertainty
relation between mean energy and inverse temperature is then
obtained (see below). A main protagonist in his treatment is
Fisher's information measure \cite{roybook,f7,pla2,pla4}.

Now, power-law distributions are ubiquitous in physics, critical
phenomena being a conspicuous example \cite{goldenfeld}. In a
statistical mechanics' context they arise quite naturally if the
information measure one maximizes (subject to appropriate
constraints) in order to arrive at the equilibrium distribution is
not Shannon's one but a generalized one. A lot of work in this
respect has been devoted to Tsallis' measure (see
\cite{t0,t00,t2,fromgibbs,t3,t4} and references therein).

 In view of the importance of these results
 it should seem appropriate to revisit the Fisher-Mandelbrot link
 by taking a closer look at non-exponential distributions of the power-law kind.
  The ensuing
results will offer novel insights into
 the meaning of non-extensivity: Fisher's measure involves all energy
 moments. Some new features of the thermal uncertainty subject will
 also be revealed.
We proceed first to a brief reminder of Fisher-related concepts.

%%%%%%%%%%%%%%%%%%%%%%%%%%%%%%%%%%%%%%%%%%%%%%%%%%%%%%%%%%%%%%%%%%%%
\section{A brief Fisher primer}
%%%%%%%%%%%%%%%%%%%%%%%%%%%%%%%%%%%%%%%%%%%%%%%%%%%%%%%%%%%%%%%%%%%%%

Estimation theory \cite{cramer} provides one with a powerful
result that needs to be  quoted before embarking into the present
discussion. Consider a system that is specified by a physical
parameter $\theta$. Let {\bf x} be a stochastic variable  and
$p_\theta({\bf x})$ the probability density for this variable,
which depends on the parameter $\theta$.  An observer makes a
measurement of
 ${\bf x}$ and
has to best infer $\theta$ from this  measurement,
 calling the
resulting estimate $\tilde \theta=\tilde \theta({\bf x})$. One
wonders how well $\theta$ can be determined. Estimation theory
asserts \cite{cramer} that the best possible estimator $\tilde
\theta({\bf x})$, after a very large number of ${\bf x}$-samples
is examined, suffers a mean-square error $e^2$ from $\theta$ that
obeys a relationship involving Fisher's $I$, namely, $Ie^2=1$,
where the Fisher information measure $I$ is of the form
\begin{equation}
I=\int \,d{\bf x}\,p_\theta({\bf x})\,\left\{\frac{\frac{\partial
p_\theta}{
\partial \theta}}{p_\theta({\bf x})}\right\}^2 =
\left\langle   \left[ \frac{1}{p_\theta({\bf x})} \frac{\partial
p_\theta}{
\partial \theta} \right]^2 \right\rangle \label{ifisher}.
\end{equation}

This ``best'' estimator is called the {\it efficient} estimator.
Any other estimator must have a larger mean-square error. The only
proviso to the above result is that all estimators be unbiased,
i.e., satisfy $ \langle \tilde \theta({\bf x}) \rangle=\,\theta
\label{unbias}$.

Thus, Fisher's information measure has a lower bound, in the sense
that, no matter what parameter of the system  we choose to
measure, $I$ has to be larger or equal than the inverse of the
mean-square error associated with  the concomitant   experiment.
This result, i.e.,

\be \label{rao} I\,e^2\,\ge \,1,
 \ee
 is referred to as the
Cramer-Rao (CR) bound, and constitutes a very powerful statistical
result \cite{roybook}. Applications of Fisher's information
measure to different physical problems have proliferated in the
last 12 years (see details and references in Frieden's book
\cite{roybook}).

%%%%%%%%%%%%%%%%%%%%%%%%%%%%%%%%%%%%%%%%%%%%%%%%%%%%%%%%%%%%%%%%%%%%%%%%%%%%%
\section{The thermal uncertainty relation}
%%%%%%%%%%%%%%%%%%%%%%%%%%%%%%%%%%%%%%%%%%%%%%%%%%%%%%%%%%%%%%%%%%%%%%%%%%%%%

Mandelbrot \cite{Mandelbrot} has established which is the form of
the probability distribution $p_\gamma (U)$ that allows for an
adequate description of the energy fluctuations of a system in
contact with a heat bath at the (inverse) temperature
$\gamma=1/T$. It is required that  estimators for $\gamma$ should
be functions of the energy $U$ only \cite{Uffink} (one demands
sufficiency of the estimator \cite{Uffink}). We are  led to the
canonical distribution
\begin{equation} \label{canon}
p_\gamma(U)=g(U)\,\frac{e^{-\gamma U}}{Z(\gamma)},
\end{equation}
with $Z(\gamma)= \int dU g(U) e^{-\gamma U}$ the partition
function and $g(U)$ the structure function, which would be
interpreted as a measure of the number of microscopic states
compatible with energy $U$ \cite{Uffink}.

Let us address the question of estimating the unknown parameter
$\gamma$ of the system by measurements of the energy. In this
case the Fisher information reads \cite{Uffink}

\be \label{fish} I(\gamma)= (\Delta_{\gamma} U)^2= \langle U^2
\rangle_{\gamma} - \left\langle U \right\rangle_{\gamma}^2 .\ee
\noindent The CR inequality for unbiased estimators $\tilde
\gamma$ then yields

\be \label{fisg2}  \Delta_{\gamma} U \Delta \tilde \gamma \ge 1,
\label{um} \ee which is Mandelbrot's uncertainty relation between
energy and temperature, expressing that the efficiency with which
temperature can be estimated is bounded by the spread in energy.
{\it This does not entail that the temperature does really
fluctuate}. It is assumed throughout that the distribution
function (\ref{canon}) with {\it fixed} $\gamma$ provides one with
an adequate description. Instead, the {\it estimators are
fluctuating, random quantities}. Their standard deviation is
employed as a criterion to indicate the quality with which the
inverse temperature is estimated \cite{Uffink}.

We can translate the preceding considerations into a {\it
microscopic}, statistical mechanics' language as follows: i) you
start with  a system in contact with a heath bath at the
temperature $T$, described by Gibbs' canonical distribution
(\ref{canon}), ii) {\it role switch}: regard the associated
inverse temperature (originally a variational Lagrange multiplier
in the entropy maximization process \cite{jaynes,katz}) as an
estimator, iii) consider the Fisher information for (\ref{canon})
together with its associated CR bound and then, iv) you get a
thermal uncertainty relation from this CR bound.

\section{Motivation for revisiting
 the thermal uncertainty derivation}

{\it  The  point we wish to make here is that  the above referred
to heath bath, employed in Mandelbrot's derivation,   cannot be a
finite one}. It is shown in \cite{fromgibbs} that if one attempts
to repeat Gibbs' celebrated derivation \cite{gibbs} for the
probability distribution (PD) that maximizes entropy for a system
in contact with a {\it finite} heath bath at the inverse
temperature $\beta$, the ensuing PD is {\it not$\,$} Gibbs
canonical distribution for the internal energy $U$. Instead, one
is forced to deal with the power-law distribution
~\cite{fromgibbs}

\be \label{detsallis} p(U) = \frac{1}{Z_q}\,\left[ 1 - \beta (1-q)
U \right]^{\frac{1}{1-q}};\,\,\,q \in \Re,\ee where $Z_q$ is a
normalization constant (the partition function). For $q=1$ the
above $q$-probability distribution becomes Gibbs' canonical one.
The distribution (\ref{detsallis}) maximizes the so-called Tsallis
information measure $S_q$, whose main feature is that of being
{\it non-extensive} if $q\ne 1$: for two independent systems
$A,\,\,B$ the entropy composition rule is \cite{t0,t00} \be
 S_q(A+B)=S_q(A)+S_q(B)+ (1-q)S_q(A)S_q(B). \ee

Remember that one of the  fundamental tenets of  information
theory is that of assigning an information content (Shannon's
measure) to any normalized probability distribution. The whole of
statistical mechanics can be elegantly re-formulated by
extremization of this measure, subject to the constraints imposed
by the {\it a priori} information one may possess concerning the
system of interest \cite{jaynes,katz}. It has been  shown in the
last decade (see, for instance \cite{t0,t00,t2,t3,t4} and
references therein) that {\it a parallel process can be undertaken
with reference to Tsallis' measure}, giving rise to what is called
non-extensive Tsallis' thermostatistics, responsible for the
successful description of an ample variety of phenomena that
cannot be explained by appeal to the conventional, {\it extensive}
one (that of Boltzmann-Gibbs)~\cite{t0,t00}.

 It is shown in \cite{fromgibbs} that a system in contact with a
{\it finite} bath is properly described by a distribution of the
type (\ref{detsallis}). The canonical distribution obtains only in
the limit in which the heath bath becomes infinite
\cite{fromgibbs}. In order to repeat the steps described in
closing the preceding Section when the protagonist is a power-law
PD, one needs to evaluate the associated Fisher measure. What
happens then with the associated, putative thermal uncertainty?
Will it remain operative? We show below that it will NOT. This
entails that the thermal uncertainty cannot be a fundamental
physical property. Our task is not a trivial one, as the content
of the following section  will show.
%%%%%%%%%%%%%%%%%%%%%%%%%%%%%%%%%%%%%%%%%%%%%%%%%%%%%%%%%%%%%%%%%%
\section{Fisher measure for  a power-law distribution}
%%%%%%%%%%%%%%%%%%%%%%%%%%%%%%%%%%%%%%%%%%%%%%%%%%%%%%%%%%%%%%%%%%
We discuss here two different information measures of the Tsallis
type, and their associated probability distributions, in order to
repeat the steps outlined previously (last paragraph of Section
III) that led to a thermal uncertainty relation for exponential
distributions. We deal first with the original Tsallis measure and
discuss afterwards the concept of escort distribution.

\subsection{Original Tsallis measure}
We start with
\begin{equation} p_\beta \big(U({\bf
x})\big) \equiv p_\beta({\bf x})  =Z_q^{-1}\,\left[ 1-(1-q)\beta
U({\bf x}) \right] ^{\frac 1{1-q}}, \label{pd}
\end{equation}
 where $\beta$ is a variational Lagrange
multiplier and $Z_q$ is the accompanying partition function
(as we sum over microstates no structure constant is needed
\cite{reif})
\begin{equation}
Z_{q}=\int d{\bf x} \left[ 1-(1-q)\beta U({\bf x}) \right] ^{\frac{1%
}{1-q}}.  \label{Zqp}
\end{equation}
\noindent We effect now the just mentioned role-switch with
regards to the meaning of the parameter $\beta$ by  introduction
of the probability distribution (\ref{pd}) into  $I$: $\beta$
plays the same role as $\gamma$ above. It becomes an estimator:
  \be I=\int d{\bf x}\,p_\beta({\bf
x})^{-1}\left[\frac{\partial p_\beta({\bf x})}{\partial \beta
}\right] ^{2}, \label{Ifisher} \ee According to Tsallis' tenets
\cite{t0} i) (Tsallis' cut-off) $$ (1-q)\beta U({\bf x}) \le
1,$$  guaranteeing non-negative probabilities and ii) one
computes mean-values according to \be \label{nonlinear}
 \langle U \rangle_q=\int d{\bf x} p_\beta^q \,U({\bf x}) .\ee

For the sake of an easier notation we shall omit, herefrom,
writing down explicitly the variable ${\bf x}$.
 We need to evaluate the integrand of Eq. (\ref{Ifisher}).
 By using (\ref{pd}) one finds
that

\be
 \frac{\partial p_\beta}{\partial \beta}=
 - p_\beta^q U Z_q^{q-1}-p_\beta Z_q^{-1}\frac{\partial Z_q}{\partial \beta}
\ee
and by (\ref{Zqp})

\be \frac{\partial Z_q}{\partial \beta}=-Z_q^q \langle U
\rangle_q. \ee As a consequence, one has, for the integrand in
(\ref{Ifisher})
 \be
p_{\beta}^{-1}\left(\frac{\partial p_\beta}{\partial
\beta}\right)^2=Z_{q}^{2(q-1)}\left[ p_\beta^{2q-1}U^2 +
  p_\beta\langle U \rangle_q^2-2 p_\beta^{q} U \langle U \rangle_q \right],
 \ee
so that, when the above relation is replaced into (\ref{Ifisher}),
 we arrive at

 \be
I=Z_{q}^{2(q-1)}\left[\langle  p_\beta^{q-1} U^2 \rangle_q-
 \langle U \rangle_q^2 \right].\label{II}
 \ee
By suitably manipulating Eq. (\ref{pd}), it is now easy to see that

\be p_\beta^{q-1}\, Z_q^{q-1}=[ 1-(1-q)\beta U]^{-1},\label{pq1}
\ee which allows one to write, for the product of the first two factors in
the first term on the right hand side above,
 \be \label{needed}
 Z_q^{2(q-1)}\, p_\beta ^{2q-1} =p_\beta \left[1-(1-q)\beta U \right]^{-2}. \ee
\noindent

Limiting  ourselves to $q-$values such that $|q|<1$, one
 can expand  the last expression into a power series in $(1-q)
 \beta$ (the convergence of the series is assured because of Tsallis' cut-off):

 \begin{equation}
Z_q^{2(q-1)}\,p_\beta ^{2q-1}=p_\beta
\sum_{n=0}^{\infty}(n+1)(1-q)^n \beta^n  U^n.\label{Qp}
 \end{equation}
Inserting this into the first term of the r.h.s of Eq. (\ref{II})
one finally gets
\begin{equation} \label{nuevo}
I=\sum_{n=0}^{\infty}(n+1)(1-q)^n\beta^n \nu_{n+2}- Z_q^{2(q-1)}\,\langle U
\rangle_q^2, \label{Iqf}
\end{equation}
where $\nu_{n+2}=\langle U^{n+2} \rangle_{q=1}$, is the $n+2-th$
order momentum of the probability  distribution \cite{cramer}.

It is time to give now to the last term of Eq. (\ref{Iqf}) the
form of a momentum expansion. We note first that

\be  \langle U \rangle_q^2\,Z_q^{2(q-1)} =[\langle Z_q^{q-1}U
 \rangle_q ]^2
\ee which, because of (\ref{pq1}) can be cast as

\be
\langle Z_q^{q-1}U
 \rangle_q=\left\langle U [ 1-(1-q)\beta U]^{-1}
 \right\rangle_{q=1}.
 \ee

Thus, a power-series expansion in  $(1-q)\beta$ plus Cauchy's
series' multiplication rule yield

$$ \langle U \rangle_q^2\,Z_q^{2(q-1)}
=\sum_{n=0}^{\infty}(1-q)^n\beta^n \sum_{k=0}^{n}\nu_{k+1}
 \nu_{n-k+1},$$
 and, finally

\be I=\sum_{n=0}^{\infty}(1-q)^n\beta^n
\left[\nu_{n+2}-\sum_{k=0}^{n}\nu_{k+1} \nu_{n-k+1}\right], \ee
that involves cross-correlation terms. We would like to establish
now an \`a la Mandelbrot uncertainty (\ref{um}). This turns out to
be  impossible! We cannot define a generalized uncertainty that
assimilates the second order $U$-momentum to $I$, because the
quantity $I$ contains a sum of terms in powers of the estimator
$\beta$. This negative result implies that thermal uncertainties are
 not operative for {\ finite} baths, only for (unphysical) infinite ones.

\subsection{A Tsallis-like measure: the ``escort'' one}

One may wonder whether the peculiar aspect of the mean values $
 \langle U \rangle_q=\int d{\bf x} p_\beta^q \,U({\bf x})$
  may not be responsible for the failure we have
just detected. We will repeat now the above steps using ordinary,
linear mean values. At this point we introduce the useful concept
of  escort probabilities (see \cite{beck} and references therein).
 One introduces the transformation
\begin{equation}
\label{escoltados} p_\beta^q({\bf x}) \rightarrow P_\beta({\bf
x}), \end{equation} with \begin{equation} P_\beta({\bf
x})=\frac{p_\beta^q({\bf x})}{\int d{\bf x}p_\beta^q({\bf x})},
\label{escoltas}
\end{equation}
$q$ being any real parameter. Here, of course, $p_\beta({\bf x})$
is given by (\ref{pd}). For $q=1$ we have $P_\beta\equiv p_\beta$
and, obviously, $P_\beta$ is normalized to unity. Our main theme
here is that any fundamental physical law must be invariant under
the above transformation.

General global quantities formed with escort distributions of
different order $q$, such as the different types of information or
mean values, will give more revealing information than those
formed with the original distribution only. Changing $q$ is indeed
a tool for scanning  the structure of the original distribution
\cite{beck}. However, basic relationships among expectation
values, like, say, Ehrenfest theorem, are invariant under the
escort transformation.

\subsubsection{Heisenberg's uncertainty relations are
invariant under (\ref{escoltados})}

 We start with usual coordinate-momentum relation
\begin{equation}
\Delta \widehat{x} \Delta \widehat{p}\geq \frac{\hbar }{2}
\end{equation}

where

\begin{equation}
(\Delta \widehat{x})^2=\langle \widehat{x}^2 \rangle - \langle
\widehat{x} \rangle^2
\end{equation}

while  a similar expression for the momentum fluctuation $\Delta \widehat{p}$.

 Expectation values of operators general $\widehat{A}$ are defined as customary
\begin{equation}
 \langle \widehat{A} \rangle=Tr(\widehat{\rho} \widehat{A}) \label{dv}
 \end{equation}
  where $\widehat{\rho}$ is, of course, the density (or statistical)
  operator.

 Under the transformation (\ref{escoltados}) we have
\begin{equation}
\widehat{\rho}\rightarrow
 \frac{\widehat{\rho}^q}{Tr(\widehat{\rho}^q)} \equiv
 \widehat{\Omega}
\end{equation}
  so that
\begin{equation}
 Tr(\widehat{\rho}
 \widehat{A}) \rightarrow
 Tr\left(\frac{\widehat{\rho}^q
 \widehat{A}}{Tr(\widehat{\rho}^q)}\right)\equiv
 Tr(\widehat{\Omega} \widehat{A}),
 \end{equation}
which entails

\begin{equation}
 \langle \widehat{A} \rangle \rightarrow  \langle \widehat{A}
 \rangle_{esc} \equiv
 Tr(\widehat{\Omega} \widehat{A}),
\end{equation}
i.e.,

\begin{equation}
\Delta \widehat{x}^{(esc)} \Delta \widehat{p}^{(esc)} \geq
\frac{\hbar }{2}
\end{equation}
with

\begin{equation}
\left(\Delta \widehat{x}^{(esc)}\right)^2=\langle \widehat{x}
\rangle_{esc}^2-\langle \widehat{x}^2 \rangle_{esc},
\end{equation}
and analogously for  $\Delta \widehat{p}^{(esc)}$.

The form of Heisenberg's principle remains invariant under
(\ref{escoltados}).

\subsubsection{Thermal uncertainty is not invariant under
(\ref{escoltados})} Di Sisto {\it et al.} have shown
\cite{disisto} that one can develop an alternative non-extensive
thermostatistics that employs an information measure
$S_{PP}=S_{PP}[P_\beta]$ which is a functional of the escort
distribution of order $q$. $S_{PP}$ depends upon the escort PD in
the same manner as Tsallis' measure depends on the original
distribution. The associated mean values are linear in the
probabilities. In our case

\be
 \langle U \rangle_{esc}=\int d{\bf x} \,P_\beta({\bf x}) \,U({\bf x}).
 \ee
\noindent
 We face now an ``escort''
Fisher's measure  \be I=\int d{\bf x}\,P_\beta({\bf
x})^{-1}\left[\frac{\partial P_\beta({\bf x})}{\partial \beta
}\right] ^{2}.  \label{Ifisher1} \ee

For our purposes we need to evaluate the integrand in
(\ref{Ifisher1}). Taking derivatives in (\ref{escoltas}) we find

\be \frac{\partial P_\beta}{\partial \beta}=q\,P_\beta\left\{
p_\beta^{-1} \frac{\partial p_\beta}{\partial \beta}- \left\langle
p_\beta^{-1}\frac{\partial p_\beta}{\partial
\beta}\right\rangle_{esc} \right\},\label{P} \ee so that, taking
derivatives in  (\ref{pd}) one is led to \be \frac{\partial
p_\beta}{\partial \beta}=-Z_q^{q-1} U p_\beta^q -p_\beta
Z_q^{-1}\frac{\partial Z_q}{\partial \beta}, \ee and, using the
result $$ \frac{\partial Z_q}{\partial \beta}=-Z_q^q\, Q\,
\langle U \rangle_{esc},$$ where we have defined $Q=\int d{\bf x}
p_\beta^q$, we get

\be \frac{\partial  p_\beta}{\partial \beta}=-Z_q^{q-1}\left[U
p_\beta^q -p_\beta Q \langle U \rangle_{esc}\right]. \ee

Replacement of this relation into  (\ref{P}) leads to

\be \frac{\partial  P_\beta}{\partial  \beta}=
 -q Z_q^{q-1} P_\beta\left\{p_\beta^{q-1}
U-\langle p_\beta^{q-1}U \rangle_{esc}\right\}, \ee
 and then to

 \be P_{\beta}^{-1}\left(\frac{\partial P_\beta}{\partial
\beta}\right)^2=q^2\,Z_q^{2(q-1)} P_\beta
 \left \{
  p_\beta^{2(q-1)} U^2+
 \langle p^{q-1}U
\rangle_{esc}^2- 2  p_\beta^{q-1}  U \langle p_\beta^{q-1}U
\rangle_{esc} \right\}. \ee

Finally, replacing this into (\ref{Ifisher1}),
 Fisher's  information measure acquires the appearance

\begin{eqnarray}
I  =q^{2}Z_q^{2(q-1)}\left\{\left\langle p_\beta^{2(q-1)}U^2
\right\rangle_{esc}- \left\langle p_\beta^{q-1} U
\right\rangle_{esc}^2\right\}.\label{IqH1}
\end{eqnarray}

Recourse to (\ref{needed}) seems to yield now a $2nd$ order
moment. However, it is not a $U$-moment  but one of the
``effective'' energy $$E=U/[1-(1-q)\beta U],$$ i.e.,

\be q^{-2}\,I  = \mu_E \equiv \langle E^2 \rangle_{esc}-
\left\langle E \right\rangle_{esc}^2,\label{Iqf1} \ee so that the
Cramer-Rao bound gives

\be \label{effec} \mu_E\, \Delta_{\beta} \ge  q^{-2}.    \ee

Appearances are deceptive, though.  The above is not an
uncertainty relation, because $\beta$ enters the two factors in
the l.h.s. Indeed, expansion into a $(1-q)
 \beta$-powers series and  Cauchy's  rule,
 where  $\nu_k=\langle U^\nu \rangle_{esc}$ is now a generalized momentum of order
 $\nu$,
 with  $\nu_{k,k-n}=\nu_{k} \nu_{n-k}$,
 gives

\be I=q^2\sum_{n=0}^{\infty}(1-q)^{n-1}\beta^{n-1}
\left[(n+1)\nu_{n}-\sum_{k=0}^{n}\nu_{k,k-n}\right], \ee which
shows again that it is not possible to get a thermal uncertainty
relation between $U$ and $\beta$.

%%%%%%%%%%%%%%%%%%%%%%%%%%%%%%%%%%%%%%%%%%%%%%%%%%%%%%%%%%%%%%%%%%%%%%%%%%%%%
\section{Conclusions}
%%%%%%%%%%%%%%%%%%%%%%%%%%%%%%%%%%%%%%%%%%%%%%%%%%%%%%%%%%%%%%%%%%%%%%%%%%%%%

As the main result of this Communication we find that it is
impossible to find a thermal uncertainty of the type (\ref{um}) if
the underlying probability distribution is not of the exponential
form. While for such exponential PDs $I$ can be assimilated to
the second moment of the energy, for non-exponential PDs (for
instance, of the Tsallis form)  the Fisher information measure
becomes a sum over all energy moments that involves all powers of
the estimator $\beta$ as well. This
 prevents us from re-obtaining the thermal uncertainty relation
of Mandelbrot's for non-exponential~PDs.

A physical interpretation of the above circumstance is connected
with the type of heath bath that helps our system to attain
thermal equilibrium. The thermal uncertainty relation {\it only
holds for systems in contact with an {\bf infinite} bath}, since
only in such a case the Gibbs canonical distribution strictly
applies. For {\it finite} baths one needs a Tsallis-canonical
distribution, as shown in detailed fashion in \cite{fromgibbs}.

Clearly, the status of the thermal uncertainty relation is thereby
affected. It can not be regarded as a fundamental property. These
facts could constitute a hopefully interesting footnote to the
ongoing controversy concerning uncertainty relations in
thermodynamics.

\section{Acknowledgement}  F. Pennini thanks financial support from
UNLP.


\begin{references}

\bibitem{ro} L.~ Rosenfeld, in {\it Ergodic theories}, edited by
P.~
Caldirola (Academic Press, NY, 1961).



\bibitem{Mandelbrot} B.~ Mandelbrot, {\it Ann. Math. Stat.} {\bf 33}, 1021 (1962);
 {\it IRE Trans. Inform. Theory} {\bf IT-2}, 190
(1956); {\it J. Math. Phys.} {\bf 5}, 164 (1964).

\bibitem{lavenda} B.~ Lavenda, {\it Int. J. Theor. Phys.} {\bf
26}, 1069 (1987); {\bf 27}, 451 (1988); {\it J. Phys. Chem. Sol.}
{\bf 49}, 685 (1988); {\it Statistical physics: a probabilistic
approach} (J. Wiley, NY, 1991).

\bibitem{Uffink} J.~ Uffink and J.~ van Lith, {\it Foundations of
Physics}  {\bf 29}, 655 (1999).


\bibitem{roybook}  B.~ R.~ Frieden, {\it Physics from Fisher information}
(Cambridge University Press, Cambridge, England, 1998).

\bibitem{promine} N.~ Bohr, {\it Collected works}, edited by J.~
Kalckar (North-Holland, Amsterdam, 1985), Vol. 6, pp. 316-330 and
376-377; A.~ Pais, {\it Niels Bohr's times in physics, philosophy,
and polity} (Clarendon Press, Oxford, 1991).



\bibitem{physto1}  C.~ Kittel, {\it Phys. Today}
(May 1988) 93.

\bibitem{physto2}  B.~ B.~ Mandelbrot, {\it Phys. Today}
(January 1989) 71.



\bibitem{f7}  B.~ R.~ Frieden and B.~ H.~ Soffer, {\it Phys. Rev. E} {\bf
52}, 2274 (1995).

\bibitem{gibbs} J.~ W.~ Gibbs, {\it Elementary principles in
statistical mechanics} (Yale University Press, 1903).

\bibitem{pla2}  A.~ R. Plastino and A. Plastino, {\it Phys. Rev. E} {\bf
54}, 4423 (1996).

%%%

\bibitem{pla4}  A.~ Plastino, A.~ R.~ Plastino, and H.~ G.~ Miller, {\it Phys.
Lett. A} {\bf 235}, 129 (1997).

\bibitem{goldenfeld} N.~ Goldenfeld, {\it Lectures on phase transitions and the
renormalization group} (Addison-Wesley, NY, 1992).



\bibitem{t0}  C.~ Tsallis, {\it Braz. J. of Phys.} {\bf 29}, 1 (1999), and
references therein.


\bibitem{t00}  A.~ Plastino and A.~ R.~ Plastino, {\it Braz. J. Phys.} {\bf 29}, 50 (1999).



\bibitem{t2}  C.~ Tsallis, {\it J. Stat. Phys.} {\bf 52}, 479 (1988).

\bibitem{fromgibbs} A.~ R.~ Plastino, A.~ Plastino, {\it Phys. Lett.
A} {\bf 193}, 251 (1994).


\bibitem{t3}  E.~M.~F. Curado and C.~ Tsallis, {\it J. Phys. A} {\bf
24}, L69 (1991); Corrigenda: {\bf 24}, 3187 (1991) and {\bf 25},
1019 (1992).

\bibitem{t4}  A.~ R.~ Plastino and A.~ Plastino, {\it Phys. Lett. A} {\bf
177}, 177 (1993).

\bibitem{cramer} H.~ Cramer, {\it Mathematical methods of statistics},
(Princeton University Press, Princeton, NJ, 1946).

\bibitem{jaynes}  E.~ T.~ Jaynes, {\it Phys. Rev.} {\bf 106}, 620 (1957); {\bf 108}, 171 (1957).

\bibitem{katz}  E.~ T.~ Jaynes in {\it Statistical Physics}, ed. W.~ K.~ Ford
(Benjamin, New York, 1963);\\ A.~ Katz, {\it Statistical
Mechanics}, (Freeman, San Francisco, 1967).



\bibitem{reif} F.~ Reif, {\it Statistical and thermal physics}
(McGraw-Hill, NY, 1965).



\bibitem{beck}  C.~ Beck and F.~ Schl\"{o}gl, {\it Thermodynamics of chaotic
systems} (Cambridge University Press, Cambridge, England, 1993).




\bibitem{disisto} R.~ P.~ Di Sisto, S.~ Martinez, R.~ B.~ Orellana,
A.~ R.~ Plastino, and A.~ Plastino, {\it Physica A} {\bf 265}, 590
(1999).


\end{references}
\end{document}